\documentclass[11pt]{amsart}
\usepackage{amsmath}
\usepackage{amsfonts}
\usepackage{amssymb}
\usepackage{enumerate}
\usepackage{graphicx}
\usepackage{hyperref}

\newtheorem{thm}{Theorem}[section]
\newtheorem{claim}[thm]{Claim}
\newtheorem{cor}[thm]{Corollary}

\newtheorem{lemma}[thm]{Lemma}
\newtheorem{prop}[thm]{Proposition}

\newcommand{\bra}[1]{\langle #1 |}
\newcommand{\ket}[1]{| #1 \rangle}
\newcommand{\braket}[2]{\langle #1 | #2 \rangle}
\newcommand{\ketbra}[2]{| #1 \rangle\langle #2 |}
\newcommand{\Tr}{\mathrm{Tr}}
\newcommand{\bb}[1]{\mathbb{#1}}
\newcommand{\cl}[1]{\mathcal{#1}}

\begin{document}

\title[Characterizing Operations Preserving Separability Measures]{Characterizing Operations Preserving Separability Measures via Linear Preserver Problems}

\author[N.~Johnston]{Nathaniel Johnston$^1$}
\address{$^1$Department of Mathematics \& Statistics, University of Guelph,
Guelph, ON, Canada N1G 2W1 \texttt{njohns01@uoguelph.ca}}

\begin{abstract}
	We use classical results from the theory of linear preserver problems to characterize operators that send the set of pure states with Schmidt rank no greater than $k$ back into itself, extending known results characterizing operators that send separable pure states to separable pure states. We also provide a new proof of an analogous statement in the multipartite setting. We use these results to develop a bipartite version of a classical result about the structure of maps that preserve rank-$1$ operators and then characterize the isometries for two families of norms that have recently been studied in quantum information theory. We see in particular that for $k \geq 2$ the operator norms induced by states with Schmidt rank $k$ are invariant only under local unitaries, the swap operator and the transpose map. However, in the $k=1$ case there is an additional isometry: the partial transpose map.
\end{abstract}

\maketitle

\section{Introduction}

The distinction between separable and entangled states is of paramount importance in quantum information theory, and entanglement leads to some of its most intriguing and counter-intuitive results. A pure quantum state (i.e., a complex unit vector) in the tensor product of two or more Hilbert spaces is said to be \emph{separable} if it is an elementary tensor, otherwise it is called \emph{entangled}. A natural generalization of the notion of separability for pure states is \emph{Schmidt rank} \cite{NC00}, which provides a rough measure of how entangled a state is -- a state has Schmidt rank $1$ if and only if it is separable.

A natural question that can be asked of the set of separable states is what operators it is preserved by. It is clear that local unitaries (i.e., unitaries acting on each of the component Hilbert spaces independently) send separable states to separable states. Additionally, a \emph{swap operator} that interchanges some of the component Hilbert spaces preserves the set of separable states. In fact, it has been shown \cite{HPSSSL06,AS10} in the bipartite case (i.e., the case when there are two component Hilbert spaces) that all operators preserving the set of separable states must be either a local unitary or the product of a local unitary and the (essentially unique) swap operator. In this work we generalize this result and investigate several related questions.

First, we examine the bipartite case in more depth and show that, for a fixed $k \geq 1$, if an operator sends the set of states with Schmidt rank at most $k$ back into itself, then it also must be either a local unitary or the product of a local unitary and the swap operator. Moreover, our proof of this generalization follows almost immediately from some known results from the theory of linear preserver problems. Second, we investigate the natural multipartite generalization and show that if an operator maps the set of (multipartite) separable states into itself then it must be the product of a swap operator (where now there are several different possible swap operators because there are more than two Hilbert spaces that can be swapped) and a local unitary. This multipartite result was originally proved in \cite{W67}, but we provide a new proof for completeness. Applications of our results will include characterizations of the isometries of certain families of norms that have recently been studied in quantum information theory \cite{JK10a,JK10b,JKPP11,CKS09,PPHH10} and a characterization of operators that preserve the geometric measure of entanglement in multipartite systems \cite{WG03}.

The remainder of this paper is organized as follows. A brief introduction to quantum information theory and the notation that we will use is provided in Section~\ref{sec:QI}. We then introduce linear preserver problems and describe two specific examples that will be of use to us in Section~\ref{sec:LPP}. In Section~\ref{sec:localUnitary} we use those linear preserver problems to generalize a known result characterizing operations that send separable pure states to separable pure states. We also prove a bipartite version of one of the linear preserver problems that was introduced.

In Section~\ref{sec:normPreserve} we use our results to characterize the isometries of two families of norms relevant to quantum information theory. We will close in Section~\ref{sec:multi} by generalizing a simplified version of one of our results to the multipartite case and characterizing operators that preserve the geometric measure of entanglement.

\section{Fundamentals of Quantum Information Theory}\label{sec:QI}

We will use $\cl{H}_n$ to denote a complex Hilbert space of (finite) dimension $n$ and $\cl{L}(\cl{H}_n,\cl{H}_m)$ the set of linear operators from $\cl{H}_n$ to $\cl{H}_m$ (for brevity, we will simply write $\cl{L}(\cl{H}_n)$ if $m = n$). Almost every Hilbert space we work with will be bipartite, meaning it is the tensor product of two smaller Hilbert spaces: $\cl{H}_m \otimes \cl{H}_n$. We will make use of bra-ket notation from quantum mechanics as follows: we will use ``kets'' $\ket{v} \in \cl{H}_n$ to represent unit (column) vectors and ``bras'' $\bra{v} := \ket{v}^*$ to represent the dual (row) vectors, where $(\cdot)^*$ represents the conjugate transpose. We will use $(\cdot)^T$ to represent the regular transpose of a vector or operator and we will write the standard basis vectors of $\cl{H}_n$ as $\big\{\ket{1},\ket{2},\ldots,\ket{n}\big\}$. We will sometimes refer to unit vectors as (pure) states. When dealing with vectors that do not have norm $1$, we will write them as $c\ket{v}$ where $c \in \mathbb{R}$ is the magnitude of the vector and $\ket{v} \in \cl{H}_n$ is a pure state specifying the direction of the vector.

A central fact in the theory of bipartite entanglement is the Schmidt decomposition theorem \cite[Chapter 2]{NC00}, which says that any pure state $\ket{v} \in \cl{H}_m \otimes \cl{H}_n$ can be written in a tensor decomposition $\ket{v} = \sum_{i=1}^k \alpha_i \ket{a_i} \otimes \ket{b_i}$, where $\big\{ \ket{a_i} \big\} \subset \cl{H}_m$ and $\big\{ \ket{b_i} \big\} \subset \cl{H}_n$ are orthonormal sets of vectors and $\{\alpha_i\}$ is a family of positive real constants. The standard proof of the Schmidt decomposition theorem proceeds by associating $\cl{H}_m \otimes \cl{H}_n$ with $\cl{L}(\cl{H}_n,\cl{H}_m)$ via the isomorphism defined on rank-$1$ tensors by $\ket{a} \otimes \ket{b} \mapsto \ket{a}\overline{\bra{b}}$ -- we will refer to this isomorphism as the ``vector-operator isomorphism''. Extending by linearity allows us to associate an arbitrary pure state $\ket{v}$ in this way to an operator $A_v \in \cl{L}(\cl{H}_n,\cl{H}_m)$. The Schmidt decomposition of $\ket{v}$ is then nothing more than the singular value decomposition of $A_v$: the coefficients $\{\alpha_i\}$ are the singular values of $A_v$, $\big\{ \ket{a_i} \big\}$ are its left singular vectors, and $\big\{ \ket{b_i} \big\}$ are (up to complex conjugation) its right singular vectors. We will make use of this isomorphism in the proofs of Theorems~\ref{thm:main} and~\ref{thm:skIso1}.

Of particular importance for us will be the notion of Schmidt rank (or tensor rank) for a pure states $\ket{v}$, which is defined as the number of non-zero terms in its Schmidt decomposition (or equivalently as the rank of the operator $A_v$ to which $\ket{v}$ is associated). We will denote the Schmidt rank of the vector $\ket{v}$ by $SR(\ket{v})$. Note that $1 \leq SR(\ket{v}) \leq \min\{m,n\}$ for all $\ket{v} \in \cl{H}_m \otimes \cl{H}_n$. Pure states with Schmidt rank equal to $1$ are called \emph{separable}, whereas pure states with Schmidt rank greater than $1$ are called \emph{entangled}.

One fundamental question that can be asked of a unitary $U \in \cl{L}(\cl{H}_m) \otimes \cl{L}(\cl{H}_n)$ is whether or not it can alter the Schmidt rank of pure states. It is not difficult to see that if $U$ is a tensor product of unitaries then it preserves Schmidt rank. Similarly, Schmidt rank is preserved by the ``swap operator'' $S \in \cl{L}(\cl{H}_n) \otimes \cl{L}(\cl{H}_n)$ defined by $S(\ket{a} \otimes \ket{b}) = \ket{b} \otimes \ket{a}$ for all $\ket{a},\ket{b} \in \cl{H}_n$. From these observations it will become important for us to consider unitaries $U \in \cl{L}(\cl{H}_m) \otimes \cl{L}(\cl{H}_n)$ of the form
		\begin{align}\label{eq:localU}
			U = W \otimes Y	\quad \text{ or } \quad n = m \text{ and } U = S(W \otimes Y),
		\end{align}
		\noindent for some unitaries $W \in \cl{L}(\cl{H}_m)$ and $Y \in \cl{L}(\cl{H}_n)$. In slightly more generality, we will consider nonsingular operators $L \in \cl{L}(\cl{H}_m) \otimes \cl{L}(\cl{H}_n)$ of the form
		\begin{align}\label{eq:localL}
			L = P \otimes Q	\quad \text{ or } \quad n = m \text{ and } L = S(P \otimes Q),
		\end{align}
		\noindent for some nonsingular $P \in \cl{L}(\cl{H}_m)$ and $Q \in \cl{L}(\cl{H}_n)$.

We will close this section with a simple lemma that shows that any operator that preserves the length of separable pure states must be unitary (and thus must preserve the length of \emph{all} pure states). We will later study the linear isometries under other norms that are related to the Schmidt rank of pure states.

\begin{lemma}\label{lem:sepUnitary}
	Let $U \in \cl{L}(\cl{H}_m) \otimes \cl{L}(\cl{H}_n)$. Then $\big\| U\ket{v} \big\| = \big\| \ket{v} \big\|$ for all separable $\ket{v} \in \cl{H}_m \otimes \cl{H}_n$ if and only if $U$ is unitary.
\end{lemma}
\begin{proof}
The ``if'' implication is trivial. To see the ``only if'' implication, note that $\bra{v}U^*U\ket{v} = 1$ for all separable $\ket{v}$, so
\begin{align}\label{eq:UUI1}
	\bra{v}(U^*U - I)\ket{v} = 0 \quad \forall \, \ket{v} \in \cl{H}_m \otimes \cl{H}_n \text{ with } SR(\ket{v}) = 1.
\end{align}

\noindent We claim that this implies $U^*U = I$ and so $U$ must be unitary. To prove this claim, write $U^*U - I = \sum_{i,j=1}^m \ketbra{i}{j} \otimes X_{i,j}$, where $\big\{ X_{i,j} \big\} \subset \cl{L}(\cl{H}_n)$. If $\ket{v} = \ket{v_1}\otimes \ket{v_2}$, then from~\eqref{eq:UUI1} we have
	\begin{align}\label{eq:UUI2}
		\bra{v_1}\Big( \sum_{i,j=1}^m \bra{v_2}X_{i,j}\ket{v_2} \ketbra{i}{j}  \Big)\ket{v_1} = 0 \quad \forall \, \ket{v_1} \in \cl{H}_m,\ket{v_2} \in \cl{H}_n.
	\end{align}
	
	\noindent It follows from Equation~\eqref{eq:UUI2} that $\sum_{i,j=1}^m \bra{v_2}X_{i,j}\ket{v_2} \ketbra{i}{j} = 0$ for all $\ket{v_2} \in \cl{H}_n$. However, because the set of operators $\big\{ \ketbra{i}{j} \big\}_{i,j=1}^m$ is linearly independent, this implies that
	\begin{align}\label{eq:UUI3}
		\bra{v_2}X_{i,j}\ket{v_2} = 0 \quad \forall \, i,j \text{ and } \forall \, \ket{v_2} \in \cl{H}_n. 
	\end{align}
	\noindent It follows that $X_{i,j} = 0$ for all $i,j$ and so $U^*U = I$.
\end{proof}

\section{Linear Preserver Problems}\label{sec:LPP}

Many of our results will stem from classical results about linear maps on complex matrices that preserve some of their properties (such as their rank, singular values, or operator norm). The problem of characterizing such maps is known as a linear preserver problem, and the interested reader can find an overview of the subject in \cite{LP01,LT92,GLS00}.

One linear preserver problem that is particularly well-known in quantum information theory is the problem of characterizing maps $\Phi : \cl{L}(\cl{H}_n) \rightarrow \cl{L}(\cl{H}_n)$ such that $(id_n \otimes \Phi)(X)$ is positive semidefinite whenever $X \in \cl{L}(\cl{H}_n) \otimes \cl{L}(\cl{H}_n)$ is positive semidefinite -- such maps are said to be \emph{completely positive}. It is well-known that $\Phi$ is completely positive if and only if there exist operators (called \emph{Kraus operators}) $\big\{A_i\big\} \subset \cl{L}(\cl{H}_n)$ such that $\Phi(X) = \sum_i A_i X A_i^*$ \cite{C75}. A completely positive map that is trace-preserving (i.e., $\Tr(\Phi(X)) = \Tr(X)$ for all $X$) is called a \emph{quantum operation}, as such maps represent the evolution of quantum states in the Schr\"{o}dinger picture of quantum dynamics. The interested reader is directed to \cite{Paulsentext} for a detailed discussion of the structure of completely positive maps.

Another linear preserver problem states that if $\Phi : \cl{L}(\cl{H}_n) \rightarrow \cl{L}(\cl{H}_n)$ preserves the Frobenius norm of all operators $X \in \cl{L}(\cl{H}_n)$, then it can be written in the form $\Phi(X) = \sum_i A_i X B_i^T$ where $\sum_i A_i \otimes B_i \in \cl{L}(\cl{H}_n) \otimes \cl{L}(\cl{H}_n)$ is a unitary operator -- a fact that follows easily from the vector-operator isomorphism and the fact that the Frobenius norm corresponds to the Euclidean norm through the isomorphism. For this reason we will refer to any linear map on matrices that preserves the Frobenius norm as \emph{unitary}.

A related result states that if $\Phi$ is a linear map that preserves the operator norm \cite{S25} (or indeed any unitarily-invariant norm on $\cl{L}(\cl{H}_n)$ that is not a multiple of the Frobenius norm~\cite{S81}), then there exist unitaries $U,V \in \cl{L}(\cl{H}_n)$ such that either $\Phi(X) = UXV$ or $\Phi(X) = UX^TV$. We will prove a similar statement for a family of norms related to the Schmidt decomposition of pure states in Section~\ref{sec:normPreserve}, though we will see that $U$ and $V$ must both be of the form~\eqref{eq:localU}.

The final linear preserver problem that will be of great use to us concerns maps that are rank-$k$-non-increasing (or equivalently, rank-$k$-preserving).

\begin{prop}\label{prop:rankKpreserver}
	Let $k,m,n$ be positive integers such that $1 \leq k < \min\{m, n\}$ and let $\Phi : \cl{L}(\cl{H}_n,\cl{H}_m) \rightarrow \cl{L}(\cl{H}_n,\cl{H}_m)$ be an invertible linear map. Then ${\rm rank}(\Phi(X)) \leq k$ whenever ${\rm rank}(X) \leq k$ if and only if there exist nonsingular $P \in \cl{L}(\cl{H}_m)$ and $Q \in \cl{L}(\cl{H}_n)$ such that $\Phi$ is of one of the following two forms:
	\begin{align}\label{eq:rankKform}
		\Phi(X) = PXQ \quad \text{ or } \quad n = m \text{ and } \Phi(X) = PX^TQ.
	\end{align}
\end{prop}

Proposition~\ref{prop:rankKpreserver} is more often stated for maps that send operators with rank $k$ to operators with exactly rank $k$ \cite{B88,BL90,L89}. The above stronger version involving operators of rank at most $k$ can be found in \cite[Section 3]{GLS00}.

\section{Operations Preserving the Schmidt Rank of Pure States}\label{sec:localUnitary}

We will now tackle the question of what operators map separable pure states to separable pure states in the bipartite setting. It is clear that any unitary of the form~\eqref{eq:localU} is an example of one such operator. In fact, it was shown in \cite{HPSSSL06,AS10} that \emph{all} such operators are of this form, though both proofs are quite long and involved. We now prove a stronger result -- that if an operator maps the set of pure states with Schmidt rank at most $k$ into itself for some $1 \leq k < \min\{m,n\}$ then it must be a unitary of the form~\eqref{eq:localU}. Moreover, our proof is quick and elementary thanks to Proposition~\ref{prop:rankKpreserver}.

\begin{thm}\label{thm:main}
	Let $L \in \cl{L}(\cl{H}_m) \otimes \cl{L}(\cl{H}_n)$ be an invertible linear operator and define $\cl{S}_k$ to be the set of scalar multiples of pure states with Schmidt rank no larger than $k$:
	\begin{align*}
		\cl{S}_k := \big\{ c\ket{w} \in \cl{H}_m \otimes \cl{H}_n : c \in \mathbb{R}, SR(\ket{w}) \leq k \big\}.
	\end{align*}
	
	\noindent Then the following are equivalent:
	\begin{enumerate}[(a)]
		\item There exists some $1 \leq k < \min\{m,n\}$ such that $L\cl{S}_k \subseteq \cl{S}_k$;
		\item $L\cl{S}_k = \cl{S}_k$ for all $1 \leq k \leq \min\{m,n\}$;
		\item $L$ is an operator of the form \eqref{eq:localL}.
	\end{enumerate}
	
	\noindent Furthermore, if $L$ sends pure states in $\cl{S}_k$ to pure states (i.e., it does not alter their norm) then $L$ is a unitary of the form~\eqref{eq:localU}.
\end{thm}

\begin{proof}
	It is straightforward to see that $(c) \Rightarrow (b) \Rightarrow (a)$, so we only prove the implication $(a) \Rightarrow (c)$.

	To this end, recall the vector-operator isomorphism that associates a pure state $\ket{v} \in \cl{H}_m \otimes \cl{H}_n$ with an operator $A_v \in \cl{L}(\cl{H}_n,\cl{H}_m)$. The same isomorphism associates the operator $L \in \cl{L}(\cl{H}_m) \otimes \cl{L}(\cl{H}_n)$ with a superoperator $\Phi_L : \cl{L}(\cl{H}_n,\cl{H}_m) \rightarrow \cl{L}(\cl{H}_n,\cl{H}_m)$. Then condition $(a)$ is equivalent to the statement that there exists some $1 \leq k < \min\{m,n\}$ such that ${\rm rank}(\Phi_L(A_v)) \leq k$ whenever ${\rm rank}(A_v) \leq k$. Proposition~\ref{prop:rankKpreserver} then says that there exist nonsingular $P \in \cl{L}(\cl{H}_m)$ and $Q \in \cl{L}(\cl{H}_n)$ such that $\Phi_L$ is of the form~\eqref{eq:rankKform}.
	
	The given form of $\Phi_L$ says, again via the vector--operator isomorphism, that either
	\begin{align*}
		L = P \otimes Q^T \quad \text{ or } \quad n = m \text{ and } L = S (Q^T \otimes P),
	\end{align*}
	
	\noindent completing the $(a) \Rightarrow (c)$ implication. The final claim is trivial -- if $L$ preserves the length of separable pure states then $P$ and $Q$ must each be unitary, so $L$ must be a unitary of the form~\eqref{eq:localU}.
\end{proof}

Notice that if $L$ preserves the length of states with Schmidt rank no greater than $k$, then we can ignore the invertibility hypothesis of the above result because Lemma~\ref{lem:sepUnitary} tells us that $L$ is necessarily unitary (and thus invertible). Also, we can rephrase Theorem~\ref{thm:main} in terms of completely positive maps that send pure states with small Schmidt rank to pure states with small Schmidt rank. The following corollary is in the same vein as \cite[Theorem 3]{FLPS10}.
\begin{cor}\label{cor:CPsep}
	Let $1 \leq k < \min\{m,n\}$ and let $\Phi : \cl{L}(\cl{H}_m) \otimes \cl{L}(\cl{H}_n) \rightarrow \cl{L}(\cl{H}_m) \otimes \cl{L}(\cl{H}_n)$ be an invertible completely positive linear map. Define $\cl{S}_k$ to be the set of scalar multiples of projections onto pure states with Schmidt rank no larger than $k$:
	\begin{align*}
		\cl{S}_k := \big\{ c\ketbra{w}{w} \in \cl{L}(\cl{H}_m) \otimes \cl{L}(\cl{H}_n) : c \in \mathbb{R}, SR(\ket{w}) \leq k \big\}.
	\end{align*}

	\noindent Then $\Phi(\cl{S}_k) \subseteq \cl{S}_k$ if and only if there exists an invertible $L \in \cl{L}(\cl{H}_m) \otimes \cl{L}(\cl{H}_n)$ of the form~\eqref{eq:localL} such that $\Phi(X) = L X L^*$. Furthermore, if $\Phi$ is trace-preserving (i.e., a quantum operation) then $L$ is a unitary of the form~\eqref{eq:localU}.
\end{cor}
\begin{proof}
	Since $\Phi$ is completely positive, there exist Kraus operators $\big\{ A_i \big\}$ such that $\Phi(X) = \sum_i A_i X A_i^*$. Consider the family of operators $\rho_{j,\ell} := \ketbra{j}{j} \otimes \ketbra{\ell}{\ell} \in \cl{S}_1$ where $1 \leq j \leq m$, $1 \leq \ell \leq n$. If we define $c_{i,h}\ket{a_{i,h}} := A_i\ket{h}$ to be the $h^{th}$ column of $A_i$, then
	\begin{align*}
		\Phi(\rho_{j,\ell}) = \sum_i A_i \rho_{j,\ell} A_i^* = \sum_i c_{i,(j-1)n+\ell}^2\ketbra{a_{i,(j-1)n+\ell}}{a_{i,(j-1)n+\ell}}.
	\end{align*}
	
	\noindent By hypothesis however, $\Phi(\rho_{j,\ell})$ has rank $1$ for all $j,\ell$. It follows that for all $i_1,i_2,h$ either
	\begin{align*}
		c_{i_1,h} = 0 \quad \text{ or } c_{i_2,h} = 0 \quad \text{ or } \quad \ket{a_{i_1,h}} = \ket{a_{i_2,h}}.
	\end{align*}
	
	\noindent In the case when $c_{i_1,h} = 0$ or $c_{i_2,h} = 0$ we can choose $\ket{a_{i_1,h}} = \ket{a_{i_2,h}}$ anyway. Thus we have that the columns of the Kraus operators $\big\{A_i\big\}$ differ only by constants -- we will now show that for a fixed pair $i_1, i_2$, each of these constants must actually be the same -- i.e., the Kraus operators $A_{i_1},A_{i_2}$ themselves differ only by constants.
	
	To this end, now consider the action of $\Phi$ on operators $\rho_{jr,\ell} := (\ket{j}+\ket{r})(\bra{j} + \bra{r}) \otimes \ketbra{\ell}{\ell} \in \cl{S}_1$. If we define $h_j := (j-1)n+\ell$ and $h_r := (r-1)n+\ell$ then
	\begin{align*}
		\Phi(\rho_{jr,\ell}) = \sum_i A_i \rho_{jr,\ell} A_i^* = \sum_i (c_{i,h_j}\ket{a_{i,h_j}} + c_{i,h_r}\ket{a_{i,h_r}})(c_{i,h_j}\bra{a_{i,h_j}} + c_{i,h_r}\bra{a_{i,h_r}}).
	\end{align*}
	
	\noindent Again, these operators must have rank $1$ for all $j,r,\ell$, so we see that there exist constants $d_{i_1,i_2,h_j,h_r}$ such that
	\begin{align}\label{eq:multiColumn}
		c_{i_1,h_j}\ket{a_{i_1,h_j}} + c_{i_1,h_r}\ket{a_{i_1,h_r}} = d_{i_1,i_2,h_j,h_r}(c_{i_2,h_j}\ket{a_{i_2,h_j}} + c_{i_2,h_r}\ket{a_{i_2,h_r}}) \quad \forall \,i_1,i_2,h_j,h_r,
	\end{align}
	
	\noindent where $h_j = (j-1)n+\ell$ and $h_r = (r-1)n+\ell$. Because we already know that $\ket{a_{i_1,h_j}} = \ket{a_{i_2,h_j}}$ and $\ket{a_{i_1,h_r}} = \ket{a_{i_2,h_r}}$, there are only two ways that Equation~\eqref{eq:multiColumn} can be satisfied. The first possibility is that $\ket{a_{i_1,h_j}}$ and $\ket{a_{i_2,h_r}}$ are co-linear -- in this case it is not difficult to see that $\Phi$ is not invertible, contradicting one of our hypotheses. The only other possibility is that $c_{i_1,h_j}\ket{a_{i_1,h_j}}=d_{i_1,i_2,h_j,h_r}c_{i_2,h_j}\ket{a_{i_2,h_j}}$ and $c_{i_1,h_r}\ket{a_{i_1,h_r}}=d_{i_1,i_2,h_j,h_r}c_{i_2,h_r}\ket{a_{i_2,h_r}}$ (i.e., the constant factor differences between the $h_j^{th}$ columns of $A_{i_1}$ and $A_{i_2}$ and between the $h_r^{th}$ columns of $A_{i_1}$ and $A_{i_2}$ are the same). This establishes that for all $i_1,i_2$ and all $1 \leq h \leq m$ there exists a constant $d_{i_1,i_2,h}$ such that $A_{i_1}(\ketbra{h}{h} \otimes I_n) = d_{i_1,i_2,h}A_{i_2}(\ketbra{h}{h} \otimes I_n)$ (i.e., the first $n$ columns of each $A_i$ are the same up to a constant, the next $n$ columns of each $A_i$ are the same up to another constant, and so on). To establish that each of these $m$ constants must be the same, apply this same reasoning again but consider instead the action of $\Phi$ on the operators $\ketbra{j}{j} \otimes (\ket{\ell}+\ket{r})(\bra{\ell} + \bra{r}) \in \cl{S}_1$.
	
	It follows that for all $i_1,i_2$ there exists a constant $d_{i_1,i_2}$ such that $A_{i_1} = d_{i_1,i_2}A_{i_2}$. Thus $\Phi$ can be written with just one Kraus operator $L$ (which is a multiple of $A_1$). The form of $L$ follows immediately from Theorem~\ref{thm:main}.
\end{proof}

In order to demonstrate that the invertibility hypotheses of Theorem~\ref{thm:main} and Corollary~\ref{cor:CPsep} are indeed required, consider the operator $L \in \cl{L}(\cl{H}_2) \otimes \cl{L}(\cl{H}_2)$ defined by $L := \ketbra{1}{1} \otimes \ketbra{1}{1} + \ketbra{1}{2} \otimes \ketbra{1}{2}$. It is clear that the range of $L$ is ${\rm span}(\ket{1}\otimes\ket{1})$ so $L\ket{v}$ is always a multiple of a separable state. However, neither $L$ nor $SL$ can be written as an elementary tensor $P \otimes Q$, even if $P$ and $Q$ are allowed to be singular.

The final result of this section is a bipartite version of Proposition~\ref{prop:rankKpreserver} in the case of rank-$1$ operators. We defer the proof of this result to the appendix, as it relies in the $k = 1$ case on Theorem~\ref{thm:multipartite} which will be introduced in Section~\ref{sec:multi}.
\begin{thm}\label{thm:sepLPP}
	Let $k,m,n$ be positive integers such that $1 \leq k < \min\{m, n\}$ and let $\Phi : \cl{L}(\cl{H}_m) \otimes \cl{L}(\cl{H}_n) \rightarrow \cl{L}(\cl{H}_m) \otimes \cl{L}(\cl{H}_n)$ be an invertible linear map. Define $\cl{S} \subseteq \cl{L}(\cl{H}_m) \otimes \cl{L}(\cl{H}_n)$ to be the set of rank-$1$ operators whose row and column space both have Schmidt rank no greater than $k$:
	\begin{align*}
		\cl{S} :=  \big\{ c\ketbra{v}{w} \in \cl{L}(\cl{H}_m) \otimes \cl{L}(\cl{H}_n) : c \in \mathbb{R}, SR(\ket{v}),SR(\ket{w}) \leq k \big\}.
	\end{align*}
	\noindent Then $\Phi(\cl{S}) \subseteq \cl{S}$ if and only if $\Phi$ can be written as a composition of one or more of the following maps:
	\begin{enumerate}[(a)]
		\item $X \mapsto LXM$, where $L,M \in \cl{L}(\cl{H}_m) \otimes \cl{L}(\cl{H}_n)$ are invertible operators of the form~\eqref{eq:localL},
		\item the transpose map $T$, and
		\item if $k = 1$, the partial transpose map $(id_m \otimes T)$.
	\end{enumerate}
\end{thm}

Although the result of Theorem~\ref{thm:sepLPP} as it is stated for rank-$1$ operators is enough for our purposes, one question that we leave open that is perhaps interesting is whether or not there exists a rank-$r$ extension of it.

\section{Operations Preserving Separability Norms and Measures}\label{sec:normPreserve}

Based on the Schmidt decomposition theorem, one can define the following family of norms for bipartite vectors $\ket{v} \in \cl{H}_m \otimes \cl{H}_n$ with Schmidt coefficients $\alpha_1 \geq \alpha_2 \geq \cdots \geq 0$ and bipartite operators $X \in \cl{L}(\cl{H}_m) \otimes \cl{L}(\cl{H}_n)$:
\begin{align}\label{norm:vector}
	\big\| \ket{v} \big\|_{s(k)} & := \sup_{\ket{w}}\big\{ |\braket{w}{v}| : SR(\ket{w}) \leq k \big\} = \sqrt{\sum_{i=1}^k \alpha_i^2},	\\	\label{norm:operator}
	\big\| X \big\|_{S(k)} & := \sup_{\ket{w},\ket{y}}\big\{ |\bra{w}X\ket{y}| : SR(\ket{w}),SR(\ket{y}) \leq k \big\}.
\end{align}

\noindent These norms have recently been studied in \cite{JK10a,JK10b,JKPP11,CKS09,PPHH10} and were shown to be related to the problem of characterizing $k$-positive linear maps and detecting bound entangled non-positive partial transpose states. Note that when $k = \min\{ m,n \}$, these norms coincide with the standard Euclidean and operator norms, respectively. The linear isometries under those two norms (that is, the linear maps that leave the norm unchanged) are both well-known -- the isometries for the Euclidean norm are exactly the unitary operators, and the isometries for the operator norm are the maps $\Phi$ that either take the form $\Phi(X) = UXV$ or $\Phi(X) = UX^TV$ for some unitary operators $U$ and $V$ \cite{S25}.

Based on Theorem~\ref{thm:main} and these known results in the $k = \min\{m,n\}$ case, one might naively expect that the isometries under the rest of the norms $\| \cdot \|_{s(k)}$ are the unitaries of the form~\eqref{eq:localU} -- we will show that this is indeed the case. Similarly, one might expect that the isometries of the rest of the norms $\|\cdot\|_{S(k)}$ are the maps of the form $\Phi(X) = UXV$ or $\Phi(X) = UX^TV$, where $U$ and $V$ are unitary operators of the form~\eqref{eq:localU}. We use Theorem~\ref{thm:sepLPP} to show that is also correct \emph{except} when $k = 1$, in which case there is an additional isometry: the partial transpose map $(id_m \otimes T)$.

\begin{thm}\label{thm:skIso1}
	Let $1 \leq k < \min\{m,n\}$ and $U \in \cl{L}(\cl{H}_m) \otimes \cl{L}(\cl{H}_n)$. Then $\big\|U\ket{v}\big\|_{s(k)} = \big\|\ket{v}\big\|_{s(k)}$ for all $\ket{v} \in \cl{H}_m \otimes \cl{H}_n$ if and only if $U$ is a unitary of the form~\eqref{eq:localU}.
\end{thm}
\begin{proof}
	The ``if'' implication is trivial. To see the ``only if'' implication, use the vector-operator isomorphism again to associate $\ket{v}$ with the operator $A_v$ and notice that $\big\|\ket{v}\big\|_{s(k)} = \sqrt{\sum_{i=1}^k \alpha_i^2}$, where $\{\alpha_i\}$ are the (ordered) singular values of $A_v$. Since this norm on $A_v$ is unitarily-invariant and not a multiple of the Frobenius norm, it follows \cite{S81} that the map $\Phi_U$ associated to $U$ through the isomorphism is of the form $\Phi_U(X) = WXY$ or $n = m$ and $\Phi_U(X) = WX^TY$ for some unitaries $W \in \cl{L}(\cl{H}_m), Y \in \cl{L}(\cl{H}_n)$. Thus we can write either $U = W \otimes Y^T$ or $U = S(Y^T \otimes W)$ (if $n = m$).
\end{proof}

Another method of proving Theorem~\ref{thm:skIso1} would be to first argue that $U$ must be unitary, then show that it must map the set of states with Schmidt rank at most $k$ back into itself, and then invoke Theorem~\ref{thm:main}. We will now use this approach to investigate the maps that preserve the modified operator norm $\|\cdot\|_{S(k)}$ for $k < \min\{m,n\}$, and we will use the same method in Section~\ref{sec:multi} to investigate operators that preserve the geometric measure of entanglement.

Although the proof of Theorem~\ref{thm:skIso1} was quite straightforward, proving the analogous result for the norms $\|\cdot\|_{S(k)}$ is slightly more technical and uses some group theory ideas presented in \cite{LT90}. Given a group $\cl{G} \subseteq \cl{L}(\cl{L}(\cl{H}_m) \otimes \cl{L}(\cl{H}_n))$, relevant terminology that we will use includes: $\cl{G}$ is said to be \emph{bounded} if there exists $K > 0$ such that $\big\|\Phi\| \leq K$ for all $\Phi \in \cl{G}$, it is said to be \emph{unitary} if $\Phi$ is unitary (i.e., it preserves the Frobenius norm) for all $\Phi \in \cl{G}$, and it is said to be \emph{irreducible} if ${\rm span}(\cl{G}) = \cl{L}(\cl{L}(\cl{H}_m) \otimes \cl{L}(\cl{H}_n))$.

\begin{thm}\label{thm:opNormIso}
	Let $1 \leq k < \min\{m,n\}$ and $\Phi : \cl{L}(\cl{H}_m) \otimes \cl{L}(\cl{H}_n) \rightarrow \cl{L}(\cl{H}_m) \otimes \cl{L}(\cl{H}_n)$. Then $\big\|\Phi(X)\big\|_{S(k)} = \big\|X\big\|_{S(k)}$ for all $X \in \cl{L}(\cl{H}_m) \otimes \cl{L}(\cl{H}_n)$ if and only if $\Phi$ can be written as a composition of one or more of the following maps:
	\begin{enumerate}[(a)]
		\item $X \mapsto UXV$, where $U$ and $V$ are unitaries of the form~\eqref{eq:localU},
		\item the transpose map $T$, and
		\item if $k = 1$, the partial transpose map $(id_m \otimes T)$.
	\end{enumerate}
\end{thm}
\begin{proof}
	Again, the ``if'' implication is trivial. For the ``only if'' implication, we first use some results from \cite{LT90,L94} to show that any map that preserves $\|\cdot\|_{S(k)}$ also preserves the Frobenius norm $\|\cdot\|_{F}$. We then show that these maps must send rank $1$ operators to rank $1$ operators, and finally we use Theorem~\ref{thm:main} to pin down the result.
	
	First, \cite[Lemmas 2.2 and 2.3]{LT90} (or equivalently, \cite[Theorem 2.3]{L94}) say that if $\cl{G}$ is a bounded group of linear operators that contains an irreducible unitary subgroup, then $\cl{G}$ itself is a unitary group. So define
	\begin{align*}
		\cl{G} := \big\{ \Phi : \cl{L}(\cl{H}_m) \otimes \cl{L}(\cl{H}_n) \rightarrow \cl{L}(\cl{H}_m) \otimes \cl{L}(\cl{H}_n) : \big\|\Phi(X)\big\|_{S(k)} = \big\|X\big\|_{S(k)} \text{ for all } X \big\}.
	\end{align*}
	
	\noindent Clearly $\cl{G}$ is bounded because it is the set of isometries under the norm $\|\cdot\|_{S(k)}$ and all norms on a finite-dimensional space are equivalent. Additionally, $\cl{G}$ contains the unitary subgroup
	\begin{align*}
		\cl{G}_S := \big\{ \Phi \in \cl{G} : \Phi(X) = (U_1 \otimes U_2)X(V_1 \otimes V_2) \text{ for some fixed unitaries } U_1,U_2,V_1,V_2 \big\}.
	\end{align*}
	
	\noindent To see that $\cl{G}_S$ is irreducible, recall that the set of unitaries $U \in \cl{L}(\cl{H}_n)$ spans all of $\cl{L}(\cl{H}_n)$, so if we fix $U_2,V_1,V_2$ then we can find maps in $\cl{G}_S$ that span the space of operators of the form
	\begin{align*}
		\Phi(X) = (A \otimes U_2)X(V_1 \otimes V_2) \text{ for some fixed arbitrary } A \text{ and unitaries } U_2,V_1,V_2.
	\end{align*}
	
	\noindent Similarly, we can obtain any map of the form $\Phi(X) = (A \otimes B)X(C \otimes D)$ in the span of $\cl{G}_S$, where $A,B,C,D$ are arbitrary. Operators of the form $A \otimes B$ span all of $\cl{L}(\cl{H}_n) \otimes \cl{L}(\cl{H}_n)$, so the span of $\cl{G}_S$ actually contains all maps of the form $\Phi(X) = EXF$ and hence all maps of the form $\Phi(X) = \sum_iE_iXF_i$. Since all linear maps can be written in this form, it follows that $\cl{G}_S$ spans the entire space of linear maps and hence is irreducible. It follows that $\cl{G}$ is a unitary group and so if $\big\|\Phi(X)\big\|_{S(k)} = \big\|X\big\|_{S(k)}$ for all $X$, then $\big\|\Phi(X)\big\|_{F} = \big\|X\big\|_{F}$ for all $X$ as well.
	
	We will now consider how a $\| \cdot \|_{S(k)}$-isometry $\Phi$ acts on rank-$1$ operators. In particular, let $\ket{v},\ket{w} \in \cl{H}_m \otimes \cl{H}_n$ with $SR(\ket{v}),SR(\ket{w}) \leq k$. Then
	\begin{align*}
		1 = \big\| \ketbra{v}{w} \big\|_{F} = \big\| \ketbra{v}{w} \big\|_{S(k)} = \big\| \Phi(\ketbra{v}{w}) \big\|_{F} = \big\| \Phi(\ketbra{v}{w}) \big\|_{S(k)}.
	\end{align*}
	
	\noindent However, $\big\|X\big\|_{S(k)} \leq \big\|X\big\| \leq \big\|X\big\|_{F}$ for all $X$ and $\big\|X\big\| = \big\|X\big\|_{F}$ if and only if $X$ has rank $1$. In this case, $\big\|X\big\|_{S(k)} = \big\|X\big\|$ if and only if there exist $\ket{x},\ket{y}$ with $SR(\ket{x}),SR(\ket{y}) \leq k$ such that $X = \ketbra{x}{y}$. Thus $\Phi(\ketbra{v}{w}) = \ketbra{x}{y}$ for some $\ket{x},\ket{y}$ with $SR(\ket{x}),SR(\ket{y}) \leq k$. Theorem~\ref{thm:sepLPP} then applies to $\Phi$ (the fact that $\Phi$ is invertible follows from it being an isometry). To finish the proof, simply note that $\Phi$ being unitary implies that the operators $L$ and $M$ of Theorem~\ref{thm:sepLPP} must be unitary.
\end{proof}

It is perhaps somewhat surprising that the group of $\|\cdot\|_{S(k)}$ isometries is slightly larger for $k = 1$ than it is for $1 < k < \min\{m,n\}$, though it was similarly observed in \cite{JK10b} that the norm $\|\cdot\|_{S(1)}$ behaves slightly differently than $\|\cdot\|_{S(k)}$ for $1 < k < \min\{m,n\}$. For example, it was shown that if $\rho$ is a Werner state \cite{W89} then either $\|\rho\|_{S(k_1)} = \|\rho\|_{S(k_2)}$ for all $k_1,k_2$ or $\|\rho\|_{S(1)} < \|\rho\|_{S(2)} = \cdots = \|\rho\|$. Similarly, there is a family of projections \cite[Section 6]{JK10a} whose $1$-norm is known, but whose $2$-norm is unknown and whose computation would resolve a decade-old conjecture of quantum information theory: that there exist non-positive partial transpose bound entangled states.

Of course, if we consider only completely positive maps (or quantum operations) $\Phi$, then Theorem~\ref{thm:opNormIso} says that $\Phi$ preserves the norm $\| \cdot \|_{S(k)}$ ($k < \min\{m,n\}$) if and only if there exists a unitary $U$ of the form~\eqref{eq:localU} such that $\Phi(X) = UXU^*$ -- a result that follows from the facts that neither the transpose nor the partial transpose maps are completely positive and the map $X \mapsto UXV$ is completely positive if and only if $V = U^*$.

\section{Multipartite Separability Preservers}\label{sec:multi}

Thus far we have only considered operations that preserve separability and entanglement in bipartite quantum systems -- i.e., Hilbert spaces that are the tensor product of two smaller Hilbert spaces. We now consider the separability preserver problem in the case of multipartite quantum systems $\cl{H}_{n_1} \otimes \cdots \otimes \cl{H}_{n_p}$ where $p \geq 3$. In particular, we will show that exactly what might naively be expected to happen in this more general setting does indeed happen -- an operator $U$ sends separable pure states to separable pure states if and only if it is a composition of local unitaries and a swap operator. The difference is that in the bipartite case there were only two subsystems to swap so there were only two possible swap operators (the identity operator and the operator $S$, which we simply referred to as \emph{the} swap operator in the bipartite setting). In the multipartite case, there are $p!$ different swap operators, each corresponding to a permutation of the $p$ subsystems.

To make this result rigorous, we must first clarify our terminology in this setting. A pure state $\ket{v} \in \cl{H}_{n_1} \otimes \cdots \otimes \cl{H}_{n_p}$ is said to be separable if it can be written in the form $\ket{v} = \ket{v_1} \otimes \cdots \otimes \ket{v_p}$, where $\ket{v_i} \in \cl{H}_{n_i}$ for all $i$. Given a permutation $\sigma : \{1,\ldots,p\} \rightarrow \{1,\ldots,p\}$, we will define the swap operator $S_\sigma : \ket{v_1} \otimes \cdots \otimes \ket{v_p} \mapsto \ket{v_{\sigma(1)}} \otimes \cdots \otimes \ket{v_{\sigma(p)}}$ to be the operator that permutes the $p$ subsystems according to $\sigma$. Additionally, we will define $\cl{S}$ to be the set of scalar multiples of \emph{multipartite} separable pure states:
	\begin{align*}
		\cl{S} := \big\{ c\ket{w} \in \cl{H}_{n_1} \otimes \cdots \otimes \cl{H}_{n_p} : c \in \mathbb{R}, \text{ $\ket{w}$ is separable } \big\}.
	\end{align*}
	
	\noindent We are now in a position to present the main result of this section. However, before proceeding we note that this result was also derived in \cite[Theorem 3.8]{W67}. We nonetheless present a full proof here for two reasons: our proof is significantly different than that provided in \cite{W67}, and many of the main results of the present paper (Theorem~\ref{thm:sepLPP}, Theorem~\ref{thm:opNormIso}, and Corollary~\ref{cor:GME}) rely on this result, so we prove it for completeness.

\begin{thm}\label{thm:multipartite}
	Let $L \in \cl{L}(\cl{H}_{n_1}) \otimes \cdots \otimes \cl{L}(\cl{H}_{n_p})$ be an invertible linear operator. Then $L\cl{S} \subseteq \cl{S}$ if and only if there exist invertible operators $P_i$ ($1 \leq i \leq p$) and a permutation $\sigma : \{1,\ldots,p\} \rightarrow \{1,\ldots,p\}$ such that $L = S_\sigma(P_1 \otimes \cdots \otimes P_p)$. Furthermore, if $L$ sends pure states in $\cl{S}$ to pure states (i.e., it does not alter their norm) then each $P_i$ is unitary.
\end{thm}

Before proving the theorem, we will need a few lemmas. Our first lemma says that the sum of two separable pure states is separable if and only if the states are either identical or differ on only one subsystem. When we say that two states $\ket{a} := \ket{a_1} \otimes \cdots \otimes \ket{a_p}$ and $\ket{b} := \ket{b_1} \otimes \cdots \otimes \ket{b_p}$ differ on the $i$th subsystem, we mean that $\ket{a_i} \nparallel \ket{b_i}$ -- in other words, $\ket{a_i}$ and $\ket{b_i}$ are not related by a complex number with modulus one. Conversely, if we write $\ket{a_i} \parallel \ket{b_i}$ then we mean that $\ket{a_i}$ and $\ket{b_i}$ are linearly dependent, so they differ by a complex number with modulus one, and we say that $\ket{a}$ and $\ket{b}$ agree on the $i$th subsystem.
\begin{lemma}\label{lem:multiHelper}
	Let $\ket{a} := \ket{a_1} \otimes \cdots \otimes \ket{a_p}, \ket{b} := \ket{b_1} \otimes \cdots \otimes \ket{b_p} \in \cl{H}_{n_1} \otimes \cdots \otimes \cl{H}_{n_p}$. Then $\ket{a} + \ket{b}$ is separable if and only if $\ket{a_i} \parallel \ket{b_i}$ for all indices $1 \leq i \leq p$ with the exception of at most one.
\end{lemma}
\begin{proof}
	The ``if'' implication of the lemma is trivial. We prove the ``only if'' implication by induction. The $p = 2$ case can be seen by using the standard bipartite vector to operator isomorphism to associate $\ket{a_1} \otimes \ket{a_2}$ and $\ket{b_1} \otimes \ket{b_2}$ with $\ket{a_1}\overline{\bra{a_2}}$ and $\ket{b_1}\overline{\bra{b_2}}$, respectively. Then $\ket{a_1} \otimes \ket{a_2} + \ket{b_1} \otimes \ket{b_2}$ is separable if and only if $\ket{a_1}\overline{\bra{a_2}} + \ket{b_1}\overline{\bra{b_2}}$ is a rank one operator. This operator is rank one if and only if $\ket{a_1} \parallel \ket{b_1}$ or $\ket{a_2} \parallel \ket{b_2}$ (or both), which establishes the base case $p = 2$.
	
	Now suppose that the claim is true for some particular $p \geq 2$ and consider linear functions $f_{i,\ket{v}} : \cl{H}_{n_1} \otimes \cdots \otimes \cl{H}_{n_{p+1}} \rightarrow \cl{H}_{n_1} \otimes \cdots \otimes \cl{H}_{n_{i-1}} \otimes \cl{H}_{n_{i+1}} \otimes \cdots \otimes \cl{H}_{n_{p+1}}$ defined on elementary tensors by
	\begin{align*}
		f_{i,\ket{v}}(\ket{a_1} \otimes \cdots \otimes \ket{a_{p+1}}) = \braket{v}{a_i} \ket{a_1} \otimes \cdots \otimes \ket{a_{i-1}} \otimes \ket{a_{i+1}} \otimes \cdots \otimes \ket{a_{p+1}}.
	\end{align*}
	
	\noindent Clearly $f_{i,\ket{v}}(\ket{a})$ is always a multiple of a separable state whenever $\ket{a}$ is separable. Now pick two vectors $\ket{a} := \ket{a_1} \otimes \cdots \otimes \ket{a_{p+1}}$, $\ket{b} := \ket{b_1} \otimes \cdots \otimes \ket{b_{p+1}}$ such that $\ket{a_i} \nparallel \ket{b_i}$ and $\ket{a_j} \nparallel \ket{b_j}$ for some $i \neq j$. Pick another index $k \neq i,j$ and choose a state $\ket{v} \in \cl{H}_{n_k}$ such that $\braket{v}{a_k},\braket{v}{b_k} \neq 0$. Then $f_{k,\ket{v}}(\ket{a})$ and $f_{k,\ket{v}}(\ket{b})$ are nonzero multiples of separable pure states living on a tensor product of $p$ Hilbert spaces that differ on the $i^{th}$ and $j^{th}$ subsystems. By the inductive hypothesis, $f_{k,\ket{v}}(\ket{a} + \ket{b})$ is not separable, so neither is $\ket{a} + \ket{b}$.
\end{proof}

\begin{lemma}\label{lem:multiHelperCor}
	Let $L \in \cl{L}(\cl{H}_{n_1}) \otimes \cdots \otimes \cl{L}(\cl{H}_{n_p})$ be an invertible operator such that $L\cl{S} \subseteq \cl{S}$. Let $1 \leq r \leq p$ and $\ket{a} := \ket{a_1} \otimes \cdots \otimes \ket{a_p}, \ket{b} := \ket{b_1} \otimes \cdots \otimes \ket{b_p} \in \cl{H}_{n_1} \otimes \cdots \otimes \cl{H}_{n_p}$ be such that there are exactly $r$ indices $h$ with $\ket{a_h} \nparallel \ket{b_h}$. If we write
	\begin{align*}
		L\ket{a} = c_a\ket{a_1^\prime} \otimes \cdots \otimes \ket{a_p^\prime} \ \ \ \ \text{ and } \ \ \ \ L\ket{b} & = c_b\ket{b_1^\prime} \otimes \cdots \otimes \ket{b_p^\prime},
	\end{align*}
	then there are at most $r$ indices $h^\prime$ with $\ket{a_{h^\prime}^\prime} \nparallel \ket{b_{h^\prime}^\prime}$.
\end{lemma}
\begin{proof}
	Suppose without loss of generality that $\ket{a}$ and $\ket{b}$ differ on the first $r$ subsystems. Then define
	\begin{align*}
		\ket{a^{(i)}} := \ket{b_1} \otimes \cdots \otimes \ket{b_{i}} \otimes \ket{a_{i+1}} \otimes \cdots \otimes \ket{a_p} \quad \text{for $0 \leq i \leq r$},
	\end{align*}
	with the understanding that $\ket{a^{(0)}} = \ket{a}$ and $\ket{a^{(r)}} = \ket{b}$. It is clear that, for any $1 \leq i \leq r$, $\ket{a^{(i-1)}} + \ket{a^{(i)}}$ is separable, so $L(\ket{a^{(i-1)}} + \ket{a^{(i)}})$ is separable as well, which implies via Lemma~\ref{lem:multiHelper} that $L\ket{a^{(i-1)}}$ and $L\ket{a^{(i)}}$ differ on at most one subsystem. It follows that $L\ket{a}$ and $L\ket{b}$ differ on at most $r$ subsystems.
\end{proof}

Our final lemma shows that if $L\cl{S} \subseteq \cl{S}$ and three separable states $\ket{v},\ket{x},\ket{y}$ are such that $\ket{x}$ and $\ket{y}$ each differ from $\ket{v}$ on a single subsystem, then $L\ket{x}$ and $L\ket{y}$ differ from $L\ket{v}$ on a single subsystem as well. Furthermore, $\ket{x}$ and $\ket{y}$ differ from $\ket{v}$ on the same subsystem if and only if $L\ket{x}$ and $L\ket{y}$ differ from $L\ket{v}$ on the same subsystem.
\begin{lemma}\label{lem:multiHelperB}
	Let $L \in \cl{L}(\cl{H}_{n_1}) \otimes \cdots \otimes \cl{L}(\cl{H}_{n_p})$ be an invertible operator such that $L\cl{S} \subseteq \cl{S}$. Let $1 \leq i,j \leq p$ and
	\begin{align*}
		\ket{v} & := \ket{v_1} \otimes \cdots \otimes \ket{v_p}, \\
		\ket{x} & := \ket{v_1} \otimes \cdots \otimes \ket{v_{i-1}} \otimes \ket{\tilde{x}} \otimes \ket{v_{i+1}} \otimes \cdots \otimes \ket{v_p} \quad (\ket{\tilde{x}} \nparallel \ket{v_i}), \\
		\ket{y} & := \ket{v_1} \otimes \cdots \otimes \ket{v_{j-1}} \otimes \ket{\tilde{y}} \otimes \ket{v_{j+1}} \otimes \cdots \otimes \ket{v_p} \quad (\ket{\tilde{y}} \nparallel \ket{v_j}).
	\end{align*}
	Write
	\begin{align*}
		L\ket{v} & = c_v\ket{v_1^\prime} \otimes \cdots \otimes \ket{v_p^\prime} \text{ for some } c_v \in \bb{C} \text{ and } \ket{v_h^\prime} \in \cl{H}_{n_h} \ (1 \leq h \leq p).
	\end{align*}
	Then there exist $1 \leq k,\ell \leq p$, $c_x,c_y \in \bb{C}$, and $\ket{\tilde{x}^\prime} \in \cl{H}_{n_k}, \ket{\tilde{y}^\prime} \in \cl{H}_{n_\ell}$ such that
	\begin{align*}
		L\ket{x} & = c_x\ket{v_1^\prime} \otimes \cdots \otimes \ket{v_{k-1}^\prime} \otimes \ket{\tilde{x}^\prime} \otimes \ket{v_{k+1}^\prime} \otimes \cdots \otimes \ket{v_p^\prime}, \text{ and}\\
		L\ket{y} & = c_y\ket{v_1^\prime} \otimes \cdots \otimes \ket{v_{\ell-1}^\prime} \otimes \ket{\tilde{y}^\prime} \otimes \ket{v_{\ell+1}^\prime} \otimes \cdots \otimes \ket{v_p^\prime}.
	\end{align*}
	Furthermore, $k = \ell$ if and only if $i = j$.
\end{lemma}
\begin{proof}
	It is clear that $\ket{v} + \ket{x}$ is separable, so $L(\ket{v} + \ket{x})$ is separable as well. It follows from Lemma~\ref{lem:multiHelper} that $L\ket{v}$ and $L\ket{x}$ differ on a single subsystem, which allows us to write $L\ket{x}$ in the form described by the lemma. The fact that $L\ket{y}$ can be written in the desired form is proved analogously. All that remains to be proved is the final claim that $k = \ell$ if and only if $i = j$.\\
	First suppose that $i = j$. Notice that $\ket{x} + \ket{y}$ is separable in this case, so $L(\ket{x} + \ket{y})$ must be separable as well. The fact that $k = \ell$ then follows from Lemma~\ref{lem:multiHelper}.
	
	Now suppose that $i \neq j$ (without loss of generality, suppose that $i < j$) -- we will prove that $k \neq \ell$ by contradiction. To this end, assume that $k = \ell$. Consider an arbitrary separable state $\ket{w}$ and three related states $\ket{w^{(i)}}, \ket{w^{(j)}}$ and $\ket{w^{(i,j)}}$, defined as follows:
	\begin{align*}
		\ket{w} & := \ket{w_1} \otimes \cdots \otimes \ket{w_p},\\
		\ket{w^{(i)}} & := \ket{v_1} \otimes \cdots \otimes \ket{v_{i-1}} \otimes \ket{w_i} \otimes \ket{v_{i+1}} \otimes \cdots \otimes \ket{v_p},\\
		\ket{w^{(j)}} & := \ket{v_1} \otimes \cdots \otimes \ket{v_{j-1}} \otimes \ket{w_j} \otimes \ket{v_{j+1}} \otimes \cdots \otimes \ket{v_p},\\
		\ket{w^{(i,j)}} & := \ket{v_1} \otimes \cdots \otimes \ket{v_{i-1}} \otimes \ket{w_i} \otimes \ket{v_{i+1}} \otimes \cdots \otimes \ket{v_{j-1}} \otimes \ket{w_j} \otimes \ket{v_{j+1}} \otimes \cdots \otimes \ket{v_p}.
	\end{align*}
	Our goal is to show that $L\ket{w}$ is contained within a nontrivial subspace of $\cl{H}_{n_1} \otimes \cdots \otimes \cl{H}_{n_p}$. Because $\ket{w}$ is an arbitrary separable state, and separable states span all of $\cl{H}_{n_1} \otimes \cdots \otimes \cl{H}_{n_p}$, this contradicts the fact that $L$ is invertible and will establish the lemma.
	
	Because $\ket{v} + \ket{w^{(i)}}$ and $\ket{x} + \ket{w^{(i)}}$ are separable, $L(\ket{v} + \ket{w^{(i)}})$ and $L(\ket{x} + \ket{w^{(i)}})$ are separable as well, and so by Lemma~\ref{lem:multiHelper} we have that $L\ket{v}$ and $L\ket{w^{(i)}}$ differ on a single subsystem, and similarly that $L\ket{x}$ and $L\ket{w^{(i)}}$ differ on a single subsystem. By invertibility of $L$ we know $\ket{v_k^\prime} \nparallel \ket{\tilde{x}^\prime}$ so it must be the case that $L\ket{v}$ and $L\ket{w^{(i)}}$ differ on the $k$th subsystem. A similar argument shows that $L\ket{v}$ and $L\ket{w^{(j)}}$ differ on the $k$th subsystem as well. Thus we can write
	\begin{align*}
		L\ket{w^{(i)}} & = c_{w^{(i)}}\ket{v_1^\prime} \otimes \cdots \otimes \ket{v_{k-1}^\prime} \otimes \ket{w_i^\prime} \otimes \ket{v_{k+1}^\prime} \otimes \cdots \otimes \ket{v_p^\prime}, \text{ and}\\
		L\ket{w^{(j)}} & = c_{w^{(j)}}\ket{v_1^\prime} \otimes \cdots \otimes \ket{v_{k-1}^\prime} \otimes \ket{w_j^\prime} \otimes \ket{v_{k+1}^\prime} \otimes \cdots \otimes \ket{v_p^\prime}.
	\end{align*}
	
	Similarly, $\ket{w^{(i)}} + \ket{w^{(i,j)}}$ and $\ket{w^{(j)}} + \ket{w^{(i,j)}}$ are separable so Lemma~\ref{lem:multiHelper} tells us that $L\ket{w^{(i)}}$ and $L\ket{w^{(i,j)}}$ differ on a single subsystem and that $L\ket{w^{(j)}}$ and $L\ket{w^{(i,j)}}$ differ on a single subsystem. Once again, this is only possible if $L\ket{w^{(i)}}$ and $L\ket{w^{(i,j)}}$ differ on the $k$th subsystem. Thus, there exists some $\ket{\tilde{w}^\prime} \in \cl{H}_{n_k}$ such that we can write
	\begin{align*}
		L\ket{w^{(i,j)}} = c_{w^{(i,j)}}\ket{v_1^\prime} \otimes \cdots \otimes \ket{v_{k-1}^\prime} \otimes \ket{\tilde{w}^\prime} \otimes \ket{v_{k+1}^\prime} \otimes \cdots \otimes \ket{v_p^\prime}.
	\end{align*}
	
	Now observe that $\ket{w}$ and $\ket{w^{(i,j)}}$ differ on at most $p-2$ subsystems, so Lemma~\ref{lem:multiHelperCor} tells us that $L\ket{w}$ and $L\ket{w^{(i,j)}}$ differ on at most $p-2$ subsystems as well. It follows that $L\ket{w}$ is contained within the set
	\begin{align*}
		\cl{T} := \big\{ c\ket{z_1} \otimes \cdots \otimes \ket{z_p} \in \cl{H}_{n_1} \otimes \cdots \otimes \cl{H}_{n_p} \ | \ \exists \, h \text{ with } \ket{z_h} = \ket{v_h^\prime} \big\}.
	\end{align*}
	Because $\ket{w}$ is an arbitrary separable state and separable states span all of $\cl{H}_{n_1} \otimes \cdots \otimes \cl{H}_{n_p}$, it follows that the range of $L$ is contained in the span of $\cl{T}$. Now let $\ket{z_h^\prime} \in \cl{H}_{n_h}$ for $1 \leq h \leq p$ be such that $\braket{z_h^\prime}{v_h^\prime} = 0$. Then clearly $(\bra{z_1^\prime} \otimes \cdots \otimes \bra{z_p^\prime})\ket{z} = 0$ for all $\ket{z} \in \cl{T}$, so $\cl{T}$ spans a strict subspace of $\cl{H}_{n_1} \otimes \cdots \otimes \cl{H}_{n_p}$. This contradicts invertibility of $L$ and completes the proof.
\end{proof}

\begin{proof}[Proof of Theorem~\ref{thm:multipartite}]
	As in the bipartite case, the ``if'' implication is trivial. For the ``only if'' implication, we prove the following claim:
	\begin{claim}
		Let $L \in \cl{L}(\cl{H}_{n_1}) \otimes \cdots \otimes \cl{L}(\cl{H}_{n_p})$ be an invertible operator such that $L\cl{S} \subseteq \cl{S}$. Fix $1 \leq i \leq p$ and vectors $\ket{v_h} \in \cl{H}_{n_h}$ ($i < h \leq p$). Then there exist a permutation $S_{\sigma}$, operators $P_h$ ($1 \leq h \leq i$), and vectors $\ket{w_h} \in \cl{H}_{n_h}$ ($i < h \leq p$) such that
		\begin{align*}
			S_{\sigma}L(\ket{v_1} \otimes \cdots \otimes \ket{v_p}) = P_1\ket{v_1} \otimes \cdots \otimes P_i\ket{v_i} \otimes \ket{w_{i+1}} \otimes \cdots \otimes \ket{w_{p}} \ \ \forall \, \ket{v_h} \in \cl{H}_{n_h} (1 \leq h \leq i).
		\end{align*}
	\end{claim}
	If we can prove the above claim then we are done, because if $i = p$ then we can use the fact that there exists a separable basis of $\cl{H}_{n_1} \otimes \cdots \otimes \cl{H}_{n_p}$ to conclude that $S_{\sigma}L = P_1 \otimes \cdots \otimes P_p$. Invertibility of each $P_j$ then follows from invertibility of $L$, and Theorem~\ref{thm:multipartite} is proved.
	
	To prove the claim, we proceed by induction on $i$. For the base case, assume $i = 1$ and fix vectors $\ket{v_h} \in \cl{H}_{n_h}$ ($2 \leq h \leq p$). Consider the $n_1$ vectors $\ket{v^{(j)}}$ ($1 \leq j \leq n_1$) defined by
	\begin{align*}
		\ket{v^{(j)}} := \ket{j} \otimes \ket{v_2} \otimes \cdots \otimes \ket{v_p}.
	\end{align*}
	By Lemma~\ref{lem:multiHelperB} we know that there exists $1 \leq k \leq p$, $\ket{w_h} \in \cl{H}_{n_h}$ ($h \neq k$), and $c_j \in \bb{R}$, $\ket{\tilde{v}^{(j)}} \in \cl{H}_{n_k}$ ($1 \leq j \leq n_i$) such that
	\begin{align*}
		L\ket{v^{(j)}} = c_j\ket{w_1} \otimes \cdots \otimes \ket{w_{k-1}} \otimes \ket{\tilde{v}^{(j)}} \otimes \ket{w_{k+1}} \otimes \cdots \otimes \ket{w_p} \quad \forall \, 1 \leq j \leq n_1.
	\end{align*}
	If $\sigma : \{1,2,\ldots,p\} \rightarrow \{1,2,\ldots,p\}$ is the permutation that swaps $1$ and $k$ then it follows by linearity of $L$ that there exists an operator $P_1 \in \cl{L}(\cl{H}_{n_1})$ such that
	\begin{align*}
		& S_{\sigma}L(\ket{v_1} \otimes \cdots \otimes \ket{v_p}) \\
		& \quad \quad \quad = P_1\ket{v_1} \otimes \ket{w_2} \otimes \cdots \otimes \ket{w_{k-1}} \otimes \ket{w_1} \otimes \ket{w_{k+1}} \otimes \cdots \otimes \ket{w_p} \quad \forall \, \ket{v_1} \in \cl{H}_{n_1}.
	\end{align*}
	We have thus proved the base case $i = 1$ of the claim.
	
	We now proceed to the inductive step. Assume that the claim holds for some specific value of $i$ and fix vectors $\ket{v_h} \in \cl{H}_{n_h}$ ($i+2 \leq h \leq p$). By the inductive hypothesis, there exist a permutation $S_{\sigma}$, operators $P_{h}$ ($1 \leq h \leq i$), and vectors $\ket{w_{h}} \in \cl{H}_{n_h}$ ($i+1 \leq h \leq p$) such that
	\begin{align}\begin{split}\label{eq:inducHyp}
		& S_{\sigma}L(\ket{z_1} \otimes \cdots \otimes \ket{z_i} \otimes \ket{1} \otimes \ket{v_{i+2}} \otimes \cdots \otimes \ket{v_p}) \\
		& \quad \quad \quad = P_{1}\ket{z_1} \otimes \cdots \otimes P_{i}\ket{z_i} \otimes \ket{w_{i+1}} \otimes \cdots \otimes \ket{w_{p}} \ \ \forall \, \ket{z_h} \in \cl{H}_{n_h} (1 \leq h \leq i).
	\end{split}\end{align}
	Fix $\ket{v_h},\ket{x_h} \in \cl{H}_{n_h}$ ($1 \leq h \leq i$) and consider the $2n_{i+1}$ vectors $\ket{v^{(j)}}$ and $\ket{x^{(j)}}$ ($1 \leq j \leq n_{i+1}$) defined by
	\begin{align*}
		\ket{v^{(j)}} & := \ket{v_1} \otimes \cdots \otimes \ket{v_{i}} \otimes \ket{j} \otimes \ket{v_{i+2}} \otimes \cdots \otimes \ket{v_p} \ \text{ and}\\
		\ket{x^{(j)}} & := \ket{x_1} \otimes \cdots \otimes \ket{x_{i}} \otimes \ket{j} \otimes \ket{v_{i+2}} \otimes \cdots \otimes \ket{v_p}.
	\end{align*}
	By Lemma~\ref{lem:multiHelperB} we know that there exists $1 \leq k \leq p$, independent of $j$, such that $S_{\sigma}L\ket{v^{(1)}}$ and $S_{\sigma}L\ket{v^{(j)}}$ differ only on the $k$th subsystem. If $k \leq i$ then we can create a vector $\ket{v^{(1)\prime}}$ that differs from $\ket{v^{(1)}}$ only on the $k$th subsystem and observe that $S_{\sigma}L\ket{v^{(1)}}$ and $S_{\sigma}L\ket{v^{(1)\prime}}$ differ on the $k$th subsystem as well. This contradicts Lemma~\ref{lem:multiHelperB}, so we see that $k \geq i + 1$.\\
	Similarly, there exists $i + 1 \leq \ell \leq p$, independent of $j$, such that $S_{\sigma}L\ket{x^{(1)}}$ and $S_{\sigma}L\ket{x^{(j)}}$ differ only on the $\ell$th subsystem. Now suppose there are $r$ indices $h$ such that $\ket{v_h} \neq \ket{x_h}$ ($1 \leq h \leq i$). Then for any $j$, $\ket{v^{(j)}}$ and $\ket{x^{(j)}}$ differ on $r$ of the first $i$ subsystems, so in particular $S_{\sigma}L\ket{v^{(1)}}$ and $S_{\sigma}L\ket{x^{(1)}}$ differ on $r$ of the first $i$ subsystems as well, by Equation~\eqref{eq:inducHyp}. However, $S_{\sigma}L\ket{v^{(1)}}$ and $S_{\sigma}L\ket{v^{(j)}}$ differ on the $k$th subsystem and $S_{\sigma}L\ket{x^{(1)}}$ and $S_{\sigma}L\ket{x^{(j)}}$ differ on the $\ell$th subsystem, $S_{\sigma}L\ket{v^{(j)}}$ and $S_{\sigma}L\ket{x^{(j)}}$ differ on $r+2$ subsystems if $k \neq \ell$, which contradicts Lemma~\ref{lem:multiHelperCor}. It follows that $k = \ell$ and furthermore that $S_{\sigma}L\ket{v^{(j)}}$ and $S_{\sigma}L\ket{x^{(j)}}$ agree on the $k$th subsystem for all $j$.
	
	Let $\tau : \{1,2,\ldots,p\} \rightarrow \{1,2,\ldots,p\}$ be the permutation that swaps $i+1$ and $k$. By using the fact that $\ket{v_h},\ket{x_h} \in \cl{H}_{n_h}$ ($1 \leq h \leq i$) were chosen arbitrarily, it follows that there exists $P_{i+1}$ (independent of $j$) such that
	\begin{align*}
		& S_{\tau}S_{\sigma}L(\ket{z_1} \otimes \cdots \otimes \ket{z_i} \otimes \ket{j} \otimes \ket{v_{i+2}} \otimes \cdots \otimes \ket{v_p}) \\
		= \ & P_{1}\ket{z_1} \otimes \cdots \otimes P_{i}\ket{z_i} \otimes P_{i+1}\ket{j} \otimes \ket{w_{i+2}} \otimes \cdots \otimes \ket{w_{k-1}} \otimes \ket{w_{i+1}} \otimes \ket{w_{k+1}} \otimes \cdots \otimes \ket{w_{p}} \\
		& \quad \quad \quad \quad \quad \quad \quad \quad \quad \quad \quad \quad \quad \quad \quad \quad \quad \quad \quad \quad \ \ \forall \, \ket{z_h} \in \cl{H}_{n_h} (1 \leq h \leq i), \ \forall \, 1 \leq j \leq n_{i+1}.
	\end{align*}
	The inductive step is completed by noting that $\{\ket{j}\}$ forms a basis for $\cl{H}_{n_{i+1}}$ and using linearity. The desired form of $L$ follows. The claim about each $P_i$ being unitary if $L$ does not alter the norm of pure states is trivial.
\end{proof}

One useful application of Theorem~\ref{thm:multipartite} is we can now characterize all operators that preserve the geometric measure of entanglement \cite{WG03}, which is defined by
\begin{align*}
	E(\ket{v}) := 1 - \sup_{\ket{w} \in \cl{S}}\big\{ |\braket{w}{v}|^2 \big\},
\end{align*}

\noindent where we note that the supremum is taken over only the pure states (i.e., vectors of length $1$) in $\cl{S}$. Observe that in the bipartite case we have $E(\ket{v}) = 1 - \big\|\ket{v}\big\|_{s(k)}^2$, so it is perhaps not surprising in light of Theorem~\ref{thm:skIso1} that the operators that preserve the geometric measure of entanglement are exactly the unitary operators of the form described by Theorem~\ref{thm:multipartite} -- a result that we will now demonstrate.

Instead of dealing with $E$ itself, it will be useful to consider the quantity $G(\ket{v}) := \sup_{\ket{w} \in \cl{S}}\big\{ |\braket{w}{v}| \big\}$ and simply note that $E(L\ket{v}) = E(\ket{v})$ for all $\ket{v}$ if and only if $G(L\ket{v}) = G(\ket{v})$ for all $\ket{v}$. It is not difficult to see that $G$ is a norm so the group of operators
\begin{align*}
	\cl{G} := \big\{ L \in \cl{L}(\cl{H}_{n_1}) \otimes \cdots \otimes \cl{L}(\cl{H}_{n_p}) : G(L\ket{v}) = G(\ket{v}) \text{ for all } \ket{v} \in \cl{H}_{n_1} \otimes \cdots \otimes \cl{H}_{n_p} \big\}
\end{align*}

\noindent is bounded. Furthermore, we can argue as in the proof of Theorem~\ref{thm:opNormIso} that the group
\begin{align*}
	\cl{G}_S := \big\{ U_1 \otimes \cdots \otimes U_p \in \cl{L}(\cl{H}_{n_1}) \otimes \cdots \otimes \cl{L}(\cl{H}_{n_p}) : U_i \in \cl{L}(\cl{H}_{n_i}) \text{ is unitary for all } i \big\}
\end{align*}

\noindent is an irreducible unitary subgroup of $\cl{G}$. It follows that $\cl{G}$ is a unitary group, so if $E(L\ket{v}) = E(\ket{v})$ for all $\ket{v}$ then $L$ must be unitary. Now using the fact that $E(\ket{v}) = 0$ if and only if $\ket{v}$ is separable, we see that $L\ket{v}$ must be separable whenever $\ket{v}$ is separable. By invoking Theorem~\ref{thm:multipartite} we have proved the following:
\begin{cor}\label{cor:GME}
	Let $U \in \cl{L}(\cl{H}_{n_1}) \otimes \cdots \otimes \cl{L}(\cl{H}_{n_p})$ be a linear operator. Then $E(U\ket{v}) = E(\ket{v})$ for all $\ket{v}$ if and only if there exist unitaries $U_i \in \cl{L}(\cl{H}_{n_i})$ ($1 \leq i \leq p$) and a swap operator $S_{\sigma} : \ket{v_1} \otimes \cdots \otimes \ket{v_p} \mapsto \ket{v_{\sigma(1)}} \otimes \cdots \otimes \ket{v_{\sigma(p)}}$ such that $U = S_\sigma(U_1 \otimes \cdots \otimes U_p)$.
\end{cor}

We leave it as an open question whether or not Theorem~\ref{thm:main} generalizes for arbitrary $k \geq 1$ to the multipartite case. That is, if $L$ is a multipartite invertible operator sending tensors with rank at most $k$ to tensors with rank at most $k$, does it follow that it must be of the form described by Theorem~\ref{thm:multipartite}?

\vspace{0.1in}

\noindent{\bf Acknowledgements.} The author is grateful to an anonymous referee who provided a significantly simplified proof of Theorem~\ref{thm:sepLPP}. Thanks are also extended to Christian Gogolin, Rajesh Pereira and Andreas Winter for asking helpful questions that led to this work. Thanks to David Kribs for constant support. The author was supported by an NSERC Canada Graduate Scholarship and the University of Guelph Brock Scholarship.

\section*{Appendix: Proof of Theorem~\ref{thm:sepLPP}}

We begin by presenting (without proof) a simple lemma that can be easily verified by the interested reader.
\begin{lemma}\label{lem:sepLPPhelp}
	Let $p \geq 3$. Suppose $X_1, X_2, \ldots, X_p$ are rank one matrices. If $X_i + X_j$ is rank one for all $i \neq j$ then $X_1 + X_2 + \cdots + X_p$ is rank one.
\end{lemma}

	We now prove Theorem~\ref{thm:sepLPP}. The ``if'' implication of the theorem is trivial to check, so we focus on the ``only if'' implication. Notice that there is an isomorphism between pure separable states $\ket{x_1} \otimes \ket{x_2} \otimes \ket{y_1} \otimes \ket{y_2} \in \cl{H}_m \otimes \cl{H}_m \otimes \cl{H}_n \otimes \cl{H}_n$ and rank one separable (not necessarily positive) operators $\ket{x_1}\overline{\bra{x_2}} \otimes \ket{y_1}\overline{\bra{y_2}} \in \cl{L}(\cl{H}_m) \otimes \cl{L}(\cl{H}_n)$. The $k = 1$ case of the result then follows by applying Theorem~\ref{thm:multipartite} and using this isomorphism -- the various swap operators $S_{\sigma}$ on $\cl{H}_m \otimes \cl{H}_m \otimes \cl{H}_n \otimes \cl{H}_n$ correspond on $\cl{L}(\cl{H}_m) \otimes \cl{L}(\cl{H}_n)$ to the transpose map, partial transpose map, and multiplication on the left and right by the swap operator $S$.

	For the case when $k \geq 2$, suppose $\Phi(\cl{S}) \subseteq \cl{S}$ where $\cl{S}$ is as defined in the statement of the theorem. We will prove the following two claims:
	\begin{enumerate}
		\item[I] $\Phi(\ketbra{v}{w})$ is rank one for all $\ket{v},\ket{w}$ with $SR(\ket{w}) \leq k$; and
		\item[II] $\Phi(\ketbra{v}{w})$ is rank one for all $\ket{v},\ket{w}$.
	\end{enumerate}
	Once Claim II is established we know that $\Phi$ must map the set of rank one matrices into itself and so the result follows by Proposition~\ref{prop:rankKpreserver} and Theorem~\ref{thm:main}.\\
	We first prove Claim I. Take any arbitrary states $\ket{v},\ket{w}$ with $SR(\ket{w}) \leq k$. Write
	\begin{align*}
		\ket{v} = \sum_{i=1}^n \alpha_i \ket{v_i} \ \ \text{ with } \alpha_i \in \bb{C}, SR(\ket{v_i}) = 1 \ \ \forall \, i.
	\end{align*}
	For any $i \neq j$, $\alpha_i \ketbra{v_i}{w} + \alpha_j \ketbra{v_j}{w} \in \cl{S}$ and so $\Phi(\alpha_i \ketbra{v_i}{w}) + \Phi(\alpha_j \ketbra{v_j}{w}) \in \cl{S}$ as well and hence it must be rank one. It follows from Lemma~\ref{lem:sepLPPhelp} that
	\begin{align*}
		\Phi(\ketbra{v}{w}) = \Phi( \alpha_1 \ketbra{v_1}{w}) + \cdots + \Phi( \alpha_n \ketbra{v_n}{w})
	\end{align*}
	is rank one as well, which establishes Claim I. Now take any arbitrary states $\ket{v},\ket{w}$ (not necessarily with Schmidt rank at most $k$) and write
	\begin{align*}
		\ket{w} = \sum_{i=1}^n \beta_i \ket{w_i} \ \ \text{ with } \beta_i \in \bb{C}, SR(\ket{w_i}) = 1 \ \ \forall \, i.
	\end{align*}
	For any $i \neq j$, $\Phi\big(\ket{v}(\overline{\beta_i}\bra{w_i} + \overline{\beta_j}\bra{w_j})\big) = \Phi(\overline{\beta_i}\ketbra{v}{w_i}) + \Phi(\overline{\beta_j}\ketbra{v}{w_j})$ is rank one by Claim I. It follows from Lemma~\ref{lem:sepLPPhelp} that
	\begin{align*}
		\Phi(\ketbra{v}{w}) = \Phi( \overline{\beta_1} \ketbra{v}{w_1}) + \cdots + \Phi( \overline{\beta_n} \ketbra{v}{w_n})
	\end{align*}
	is rank one as well. Claim II follows and the proof is complete.

\end{document}